\documentclass[aps,prl,float,twocolumn,a4paper,amsmath,amssymb]{revtex4}
\usepackage[dvips]{color}
\definecolor{g-blue}{rgb}{0.83,0.95,1}
\definecolor{g-yellow}{rgb}{1,1,0.7}
\definecolor{g-green}{rgb}{0.9,1,0.9}
\definecolor{green}{rgb}{0,0.6,0}
\definecolor{cyan}{rgb}{0,0.7,0.7}
\definecolor{black}{rgb}{0,0,0}
\definecolor{grey}{rgb}{0.4 ,0.4 ,0.4 }

\def\blue#1{\textcolor{blue}{#1}}

\usepackage{graphicx}
\usepackage{amsmath,bm,epsfig}
\usepackage{verbatim}
 \usepackage[normalem]{ulem}
\def \ed {\end{document}}
\def\Fbox#1{\vskip1ex\hbox to 8.5cm{\hfil\fboxsep0.3cm\fbox{%
\parbox{8.0cm}{#1}}\hfil}\vskip1ex\noindent} 
\def\be{\begin{equation}}\def\ee{\end{equation}}
\def\bea{\begin{eqnarray}}\def\eea{\end{eqnarray}}
\def\bse{\begin{subequations}}\def\ese{\end{subequations}}
\newcommand{\BE}[1]{\begin{equation}\label{#1}}
\newcommand{\BEA}[1]{\begin{eqnarray}\label{#1}}
\newcommand{\BSE}[1]{\begin{subequations}\label{#1}}

\let \= \equiv \let\*\cdot \let\~\widetilde \let\^\widehat \let\-\overline

 \def\1{\bm1} 
\def\<{\left\langle} \def\>{\right\rangle}
\def\({\left(} \def\){\right)}
\def \[ {\left [} \def \] {\right ]}

\newcommand{\B}[1]{{\bm{#1}}}
\newcommand{\C}[1]{{\mathcal{#1}}} 
\newcommand{\BC}[1]{\bm{\mathcal{#1}}}
\newcommand{\Sp}[1]{^{^{\text {#1}}}} 

\begin{document}
\title{Analytic solution of the dynamics of quantum vortex reconnection}
\author{Laurent Bou\'e, Dmytro Khomenko, Victor S. L'vov, Itamar Procaccia}
\affiliation{Department of Chemical Physics, Weizmann Institute of Science, Rehovot 76100, Israel}

\begin{abstract}
Experimental and simulational studies of the dynamics of vortex reconnections in quantum fluids showed
that the distance $d$ between the reconnecting vortices is close to a universal time dependence $d=D[\kappa|t_0-t|]^\alpha$
with $\alpha$ fluctuating around 1/2 and $\kappa=h/m$ is the quantum of circulation. Dimensional analysis, based on the
assumption that the quantum of circulation $\kappa=h/m$ is the only relevant parameter in the problem, predicts
$\alpha=1/2$. The theoretical calculation of the dimensionless coefficient $D$ in this formula remained an open problem. In this Letter we present an analytic calculation of $D$ in terms of the given geometry of the reconnecting vortices. We start from the numerically observed generic geometry on the way to vortex reconnection and demonstrate that the dynamics
is well described by a self-similar analytic solution which provides the wanted information.

\end{abstract}
\date{\today}
\maketitle

The study of vortex reconnections in quantum fluids received a huge boost by the experimental visualization of this
process in liquid Helium \cite{10PFL}. Of special interest is the dynamics of the reconnecting vortices as displayed
by the minimal distance between them, denoted as $d(t)$. When $d(t)$ is much larger than the vortex core size one can
neglect the core size; then dimensional considerations based on the assumption that the quantum circulation $\kappa=h/m$ (with $h$ being Plank's constant and $m$ the mass of the $^4$He atom) is the only relevant parameter of the problem predict that \cite{85Sig,85Sch}
\begin{equation}
d(t)=D[\kappa|t_0-t|]^{\alpha} \ , \quad \alpha=1/2 \,,
\label{doft}
\end{equation}
where $D$ is a dimensionless parameter.
Experiments and simulations exhibited a range of exponents near $\alpha=1/2$ \cite{85Sig,10PFL}, but the question of the
coefficient $D$ and how to compute it theoretically remained an open question. The aim of this Letter is to solve this
open question.
\begin{figure}
 \includegraphics[width=1.1 \linewidth]{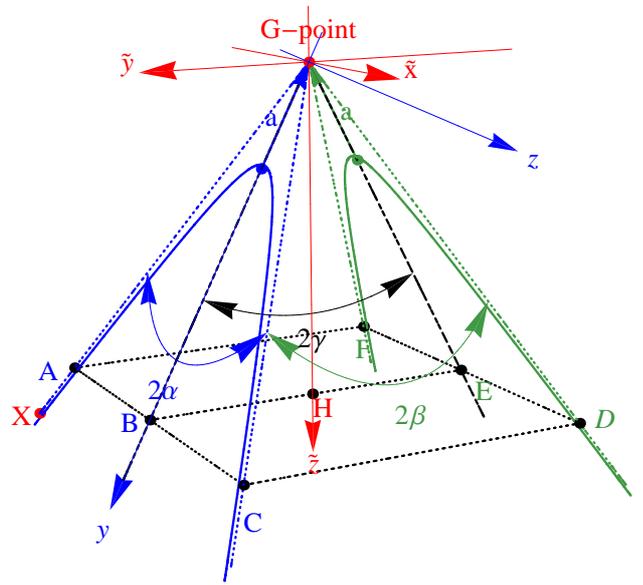}
 \caption{\label{f:1} Color online. The proposed ``generic" pyramidal configuration of two vortex lines on the way to collision and reconnection \cite{Waele}. The two vortex lines (blue and green) are hyperbolas on the two opposite side edges ACG and DFG of the ACDFG-pyramid.  The conjecture of Ref. \cite{Waele} is that the vortex lines proceed to collide at the G-point. In this example $ \alpha=60^\circ$, $ \beta=12.5^\circ$. We show in this letter that some serious modifications are necessary to reach a true self similar solution.}
\end{figure}

We first note that the core size $\delta$ in superfluid $^4$He is about one Angstr\"om, $\delta\approx 10^{-8}$ cm.
We can assume that $D(t)\gg \delta$, and describe the evolution
of quantized vortex lines by the Biot-Savart equation according to which each point of the vortex
line is swept by the velocity field produced by the other existing vortices
\begin{equation}
\B V(\B s) =\frac{\kappa}{4\pi}\int_C \frac{(\tilde {\B s}-\B s)\times d \tilde {\B s}}{|\tilde {\B s}-\B s|^3} \ .
\label{BSE}
\end{equation}
Here the vortex line is represented in a parametric form
$\B s(\xi,t)$, where $\xi$ is an arc-length, $t$ is the time and the integral
is taken over the entire vortex tangle configuration. The core size $\delta$ appears implicitly in this equation
as the cutoff length that protects the logarithmic singularity that is embodied in Eq. (\ref{BSE}). Finally the core
radius $\delta$ appears in the present problem as a logarithmic term of the form $\Lambda = \ln (d(t)/\delta)$.

The strategy employed here is to start from the remarkable numerical discovery that independently of initial conditions, near to a reconnection point vortices arrange themselves to become antiparallel, forming an evolving structure that appears like a self-similar solution ~\cite{Waele,11TYB}. In this Letter we construct an approximate self similar evolution of the vortex lines towards their reconnection, ignoring the weak logarithmic dependence on $\delta$. Thus our starting point is the vortex configuration found in~\cite{Waele} and reproduced in Fig. \ref{f:1} as pyramidal construction with A-, B-, C-, E- and F-points on the base and the G-point  on the top. Two vortices occupy two side edges as shown by the solid (A-C blue and D-F green) lines. The vortex lines are separated (at a given moment of time $t$) from the reconnection $G$-point by a distance $a(t)\to 0$ at $t\to t_0$. Far away from the G-point the lines are almost straight, approaching four semi-infinite straight   lines shown in Fig. \ref{f:1} as dashed AG- and GC-lines (blue vortex) and DG- with GH-lines (green vortex).

 In our analysis we will use two coordinate systems. The first (basic)
 $( \tilde {\B x},\tilde{\B y},\tilde{\B z})$-coordinate system (shown in red) has an origin at the G-point and  the $\tilde {\B x}$ axis is parallel to the straight ABC-  and DEF-straight lines.
  By $2 \alpha$ we denote the angle between AG- and  CG-lines, equal to the angle between DG- and  FG-lines.  By $2\beta$ we denote the angle between   CG- and  DG-lines, equal to the angle between  AG- and FG-lines.  $\tilde {\B z}$ axis is directed down from G- to the point H and the $\tilde {\B y}$ axis is parallel to the BE-line. In addition we will use a (blue) $(\B x,\B  y,\B z)$-coordinate system, in which $\B x=\tilde{\B x}$, the $(\B x,\B y) $-plane coincides
 with the ACG-plane such that the $\B y$-axis is obtained by turning the $\tilde{\B y}$  by an angle $\displaystyle \frac \pi 2- \gamma$ around the $\B x$-axis. Here $2\gamma$ is the angle between the ACG- and the DEG-planes (or between BG- and EG-lines), related to $\alpha$ and $\beta$ as follows: $\cos \alpha \sin \gamma = \sin \beta$.

 The construction of the self-similar evolution will be achieved in three steps. In the first step we will disregard the tip region and approximate the vortex configuration by the four straight lines: AG, CG, DG and FG. Then we compute the velocity induced on a given vortex line, e.g. AG, by the other three.  Clearly the contribution from one straight vortex line on itself is zero, since a straight vortex line is an obvious null solution of Eq.~(\ref{BSE}).
For this goal we recall  the fundamental result of the Biot-Savart equation for the velocity field $\B V(r,\varphi)$ produced by a semi-infinite  vortex line
\begin{equation}\label{3}
 \B V(r,\varphi)=\frac{\kappa}{4 \pi r} (\cos \varphi  +1)\ .
 \end{equation}
 Here $r$ is a distance from the $\B X$-point \blue(shown in Fig.~\ref{f:1}) where the velocity is measured to the semi infinite vortex line; $\varphi$ is the angle between the semi-infinite vortex line and the line between the end point of the vortex line and the measurement point $O$. Using Eq.~\eqref{3} we can find the velocity $\B V(\B X)$  induced by the CG-, DG- and FG-lines on the $\B X$-point. A straightforward but quite cumbersome calculation that involves lots of trigonometry yields:
\begin{subequations}\label{SL}
\begin{eqnarray}\label{SLa}
\B {V\Sp {BS}}(\B X)&=&\frac{\kappa}{4\pi x}\BC V\,,\quad \BC V=\{\C V_\perp, \C V_+, 0\}\,, \\ \label{SLb}
\C V_\perp&=&-\cos  \alpha+\frac{\sin  \alpha\tan  \alpha}{\sin^2
  \alpha+\sin^2 \beta}\,, \\ \label{SLc}
\C V_+&=&\frac{\cos \gamma\sin^3  \alpha}{(\sin^2 \alpha +\sin^2 \beta)\sin  \beta}\ .
\end{eqnarray}\end{subequations}
Here $ \BC V$ is a vector of dimensionless numbers defined entirely by the geometry:     $\C V_\perp$  is the component orthogonal to the ACG-plane,   $\C V_+$ is an in-plane component, normal to the AG-line and $\C V_\parallel =0$ is the component along the AG-line.
The crucial observation of the first step of the calculation is {\em that for large $x$ the velocity field decays like $1/x$}. This asymptotic statement is independent of the near-tip structure, and will
be reproduced by the solution discusses below.

In the second step of the calculation we will adopt for a start the proposition of Ref.~\cite{Waele} that the reconnecting vortex lines have a form of two identical hyperbolae AC (blue) and FD (green) lying in the corresponding planes, see  Fig.~\ref{f:1}. The AC-hyperbola in the $(x,y)$ plane is
\begin{subequations}\label{6}
\begin{equation}\label{6A}
y_0^2(x,a(t))= a^2(t)+x^2 \cot^2  \alpha \ .
\end{equation}
We will see below that this proposition is too restrictive and it cannot be satisfied accurately by the a self-similar
solution. We thus go beyond the guess of Ref.~\cite{Waele}; we find it advantageous to use a less restrictive form which we refer to as ``quasi-hyperbola" and write as
\begin{equation}\label{sol1}
y_1^2(x,a(t),\varepsilon)= y_0^2(x,a(t))+ \frac{ \varepsilon\, a^2(t)x^2 \cot^2  \alpha}{ y_0^2(x,a(t))} \ .
\end{equation}
\end{subequations}
Clearly, $y_1(x,a(t))\to y_0(x,t)$ for $\varepsilon\to 0$ and/or $x\to \infty$, while for $x\lesssim a$ the curvature of $y_1(x,t)$ depends on $\varepsilon$.

The $y(x,t)$ lines~\eqref{6} are translated in time due to the $t$-dependence of $a(t)$ such that $a(t)\to 0$ for $t\to t_0$.   Naturally there are infinitely many mappings of a given $y(x,t)$-line to a future $y(x,t+\delta t)$-line. We remember however that these are quantized vortex lines and therefore vortex stretching does not effect the circulation. Therefore the tangential component of the velocity is not relevant for the present construction. We thus seek
self-similar kinematics by requiring that each point of the vortex line should move perpendicularly to the vortex line with
the in-plane,  normal to the line velocity $V_+$ which is:
\begin{equation}
V_+^{M}(x,t)=  \frac{da}{dt} \Big(\frac{\partial y}{\partial a }\Big )_x \Bigg / \sqrt{1+\Big(\frac{\partial y}{\partial x}\Big ) ^2_a } \ .
\label{vn}
\end{equation}
From Eq.~\eqref{vn} we see that the asymptomatic behavior of $V_+\Sp M(x,t)$ for $x\gg a(t)$  is determined by $(d{y}/da)_x$. But in step 1 we found that the asymptotic is $1/x$.  Consistency requires the following condition on the function ${y}$:
 $\displaystyle \lim_{x\to\infty}\frac{d{y}}{da} \propto \frac 1 x \ . $
It is obvious that both lines~\eqref{6} indeed satisfy this condition. In addition,
a direct calculation using Eqs. \eqref{vn} and \eqref{6} yields

\begin{widetext}
 \begin{equation}
  V_+\Sp M(x,a,\alpha,\varepsilon)=\frac{\dfrac{da}{dt}a[y_0^4(x,a)+\varepsilon  x^4 \cot ^4\alpha]  }{
   \sqrt{y_0^8(x,a)(a^2+x^2\cot^2\alpha\csc^2\alpha)+ \varepsilon x^2a^2\cot^2\alpha y_0^4(x,a)[y_0^2(x,a)+2 a^2 \cot^2 \alpha ]+\varepsilon^2 a^8x^2\cot^4\alpha}}\ . \label{Vnnew}
\end{equation}
\end{widetext}
Of course, the leading asymptotics of $V_+\Sp M$ (for $x\gg a$) must be consistent with Eqs. (\ref{SL}) which resulted from t Eq.~\eqref{BSE}. The conditions for consistency are:
\begin{equation}
\frac{da^2}{dt}=-\frac{\kappa \C V_+\cot\alpha}{2\pi\sin \alpha (1+\varepsilon) }\ . \label{conds}
\end{equation}
Solving this equation one finds $a^2(t) = A^2 \kappa(t_0-t)$, where
 \begin{equation}
A^2=\frac{\sqrt{\cos(2 \alpha)+\cos(2 \beta)}\sin  \alpha}{2 \sqrt{2}\pi
(\sin^2\alpha +\sin^2 \beta) \sin  \beta (1+\varepsilon)} \ . \label{result2}
 \end{equation}

\begin{figure}
\includegraphics[width=0.8 \linewidth]{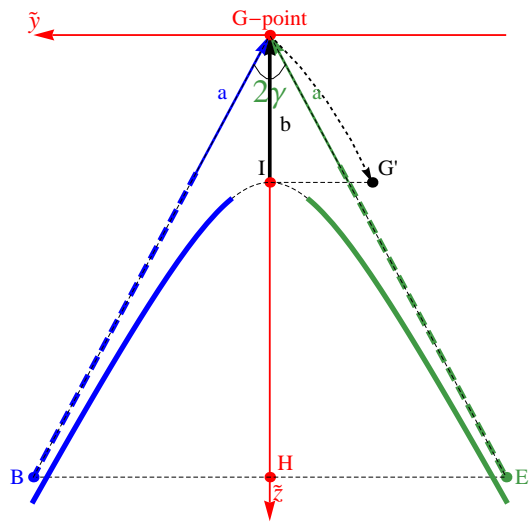}
\caption{ The third step of the analysis including the second modification to Ref.~\cite{Waele}: now the quasi-hyperbolas are embedded in a surface which is also a quasi-hyperbola. Consistency with the asymptotic solution (large $X$) requires
moving the collision G-point to a new G'-point. Note that the two vortex lines are here but only one branch in each
is shown.}
\label{bending}
\end{figure}

 Although we showed that the self-similar solutions~\eqref{6} are asymptotically consistent with the Biot-Savart result for the normal component of the velocity in the plane, we should note that the Biot-Savart result contains also a $V_\perp \Sp{BS}$ component of the velocity, orthogonal to the plane. In the considered geometry this component vanishes, $V_\perp \Sp M=0$, because the lines~\eqref{6} are moving in the plane.  Such a component would obviously destroy the self similarity of the chosen configuration of two flat quasi-hyperbolae lying each in its own plane. This is another point where we have to go beyond the guesses solution of Ref. \cite{Waele}. In the third step of the analysis we fix this discrepancy by choosing quasi-hyperbolae that lie {\em not on the planes} but on quasi-hyperbolic surfaces as shown in Fig.~\ref{bending}. Explicitly, we choose the quasi-hyperbolic surface using the relation:
\begin{subequations}\label{sol2}
\begin{equation}\label{new-h}
\tilde{z}(\tilde y,t)=y_1(\tilde y,b(t),\tilde\varepsilon)\,,
\end{equation}
with the same quasi-hyperbola~\eqref{sol1} but defined in the different(tilde) coordinate-system, see Figs.~\ref{f:1} and~\ref{bending}.

Such a construction automatically provides the desirable asymptotics because the $V_\perp\Sp M$ component of the velocity is defined now by the same type of expression as Eq.~(\ref{Vnnew}), replacing   $\alpha \Rightarrow\gamma$ and $a \Rightarrow b$:
\begin{equation}
 V_\perp\Sp M=V_+\Sp M(\tilde y,b,\gamma,\tilde{\varepsilon}) \ .
\end{equation}\end{subequations}
In the new coordinates we have asymptotically
$
\lim_{x\to\infty} \tilde{y} =x \cot  \alpha\sin \gamma   $.
In its turn, the $V_\perp\Sp M$-component of velocity will be asymptotically
\begin{equation}
\lim_{x\to\infty}  V_\perp\Sp M  =\frac{b^2 (1+\tilde{\varepsilon})}{a x\cot(\gamma)\cot  \alpha }\frac{da}{dt} \ .
\end{equation}
Now the distance $b$ is defined by the condition
 \begin{eqnarray}
\lim_{x\to\infty}  V_\perp\Sp M   = V_\perp\Sp {BS}\,,  \quad
\frac{b^2 (1+\tilde{\varepsilon})}{a x\cot(\gamma)\cot  \alpha }\frac{da}{dt} = \frac{\kappa \C V_\perp}{4\pi x}\ .
\end{eqnarray}
 After a direct calculation we find
\begin{equation}\label{b}
\frac{b}{a} =\sqrt{\frac{1+\varepsilon}{1+\tilde{\varepsilon}}\left(\sin^2 \alpha -\cot^2  \alpha\sin^2  \beta\right)}\ .
\end{equation}

Finally we note that we defined   quasi-hyperbolic surface [instead of two (blue and green) planes to embed the vortex lines in them, but did not specify
explicitly how this is done. To achieve correct embedding we employ the condition that far from the tip the vortex lines
and the measurement point $\B X$ should coincide for a large value of $x$. This condition will move the point $G$ to a new point $G'$, see Fig. \ref{bending}. The Distance $IG'$ between the collision point and the point $G'$ is determined by
the requirements that the arc length in the bent plane (quasi-hyperbolic surface) coincides with the distance in the flat plane:
\begin{equation}
IG'=\lim_{\tilde y\to \infty}\left[\frac{\tilde y}{\sin(\gamma)}-L(\tilde y)\right],
\end{equation}
where $L$ is the arc length of the quasi-hyperbola which is defined by:
\begin{equation}
 L(\tilde y)=\int\limits _0^{\tilde y} \sqrt{1+\Big(\dfrac{d\tilde{z}}{d\tilde{y'}}\Big)^2}d\tilde{y'} \label{L}\ .
\end{equation}
Now we can write the parametric equation for the configuration of vortex line in the tilde coordinate system:
 \begin{equation}\label{sol4}
\tilde{x}(x)=x\,, \quad
\tilde{y}(x)=L^{-1}[y_1(x,a,\varepsilon)-IG'] \,,
\end{equation}
where $L^{-1}$ is inverse function of $L$, defined in Eq.~\eqref{L}, and
$\tilde{z}(x)=y_1(\tilde{y}(x),b,\tilde\varepsilon)$.

\begin{figure}
\includegraphics[width=1\linewidth]{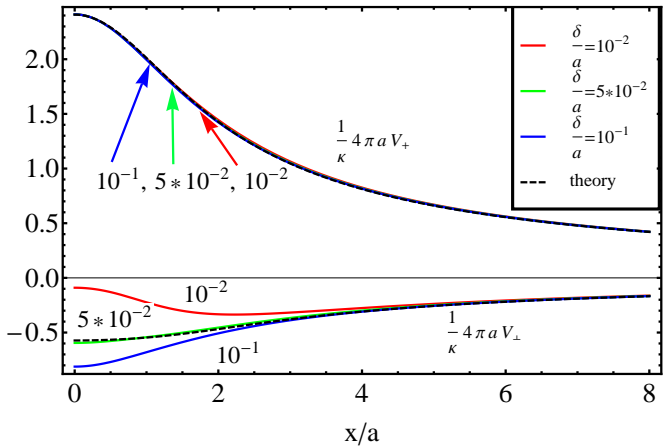}
\caption{\label{fvn} Color online. Comparison of the the normalized vortex velocity profiles  $4 a(t)\pi V_+ (x,t)/\kappa$ (in plane) and  $4 a(t)\pi V_\perp (x,t)/\kappa$ (normal to the plane), computed from the Biot-Savart equation (solid lines) and different value of $\delta/a$ with the model prediction, shown by a dashed black line.   The parameters used in the theory are $\alpha=60^\circ$, $\beta=12.5^\circ$, $\epsilon=-0.06$ and $\tilde \epsilon=1$}
\end{figure}

This finalize the self-similar solution, given by Eqs.~(\ref{sol1}, \ref{sol2},\ref{b}). While we expect this solution to be quite accurate, it cannot be exact, since
we neglected the logarithmic correction involved with the inner cutoff of the Biot-Savart integral \eqref{BSE}. To
asses the accuracy we compare the the vortex velocity found from the time-evolution of the suggested self-similar solution~(\ref{sol1}, \ref{sol2},\ref{b},\ref{sol4}), denoted as ``theory" in Fig.~\ref{fvn},
with direct numerical calculations of  the vortex velocity from  the Biot-Savart Eq.~\eqref{BSE} with the vortex configuration~(\ref{sol1}, \ref{sol2},\ref{b},\ref{sol4}). The calculations were performed at different values of $\delta/a$. The
results are presented in Fig.~\ref{fvn} as a function of $x/a$.
 \begin{figure}[t]
 \vskip 0.5cm
\includegraphics[width=1\linewidth]{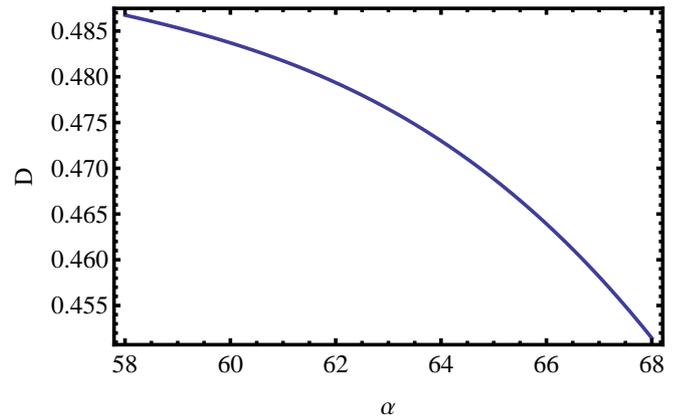}
\caption{\label{D} The coefficient D as a function of the angle $\alpha$ \blue{(in $^\circ$)} in the range of angles considered in Ref. \cite{Waele} \blue{with $\beta=12.5^\circ$}.  }
\end{figure}
We conclude that the Biot-Savart velocity always agrees with the theory far from the tip (for $x>4a$). The agreement is almost perfect for the $V_+$ component for all values of $x/a$ and $\delta/a$. On the other hand the $V_\perp$-component is
 more sensitive near the tip to the local contributions that exist in the Biot-Savart integral. It differs from the self-similar solution in the tip region and obviously cannot be fitted for all values of $\delta/a$. Nevertheless, even near to the tip the agreement is still reasonable, especially taking into account the fact that $V_\perp$ is less than a 1/3 of $V_+$ for all values of $\delta/a$.

 Finally, we return to the question of computing the coefficient $D$ in Eq. (\ref{doft}). The numerical
 simulations in Ref.~\cite{Waele} were used to determine an approximate value of $D$ in the vicinity of 0.4. In light of our analysis
 it is obvious that $D$ depends on the angle $\alpha$ and is a function rather than a number, as is indeed found
 in the experimental work, cf. \cite{10PFL}. Since Ref. \cite{Waele} estimated the value of $D$ only in the range
 $58^\circ \le \alpha \le 68^\circ$ we calculate $D$ in the same range (with the angle $\beta=12.5^\circ$). The result
is shown in Fig. \ref{D}. It is clear that the analytic prediction is in a quite close agreement with the numerical
estimate.

In summary, we presented an analytic self-similar solution for the dynamics of vortex lines approaching a reconnection.
The solution is in good agreement with  numerical simulations. In future work one needs to understand how to improve
the solution to include the reconnection event itself and the change in topology. There are reasons to believe that this
process includes the release of energy in the form of sound waves \cite{Barenghi}, leading to an asymmetry between the dynamics before
and after the reconnection. This cannot be done with Biot-Savart dynamics and calls for a fully quantum mechanical model.

\end{document}